# Nucleosynthesis in Core Collapse Supernovae

Marco Limongi[*] and Alessandro Chieffi[¶]

[*]INAF – Osservatorio Astronomico di Roma, Via Frascati 33, I-00040, Monteporzio Catone, Roma, Italy. Email: marco@oa-roma.inaf.it
[¶]INAF – Istituto di Astrofisica Spaziale e Fisica Cosmica, Via Fosso del Cavaliere, I-00133, Roma, Ital. Email: achieffi@rm.iasf.cnr.it

**Abstract.** We present the basic properties of the yields of our latest set of presupernova evolution and explosive nucleosynthesis of massive stars in the range between 11 and 120 $M_\odot$ having solar and zero metallicity.

## INTRODUCTION

Massive stars, those massive enough to explode as supernovae, play a key role in many fields of astrophysics. They are crucial in determining the evolution of the galaxies because: 1) they light up regions of stellar birth and hence induce star formation; 2) they are responsible for the production of most of the elements (among which those necessary to life); 3) they may induce mixing of the interstellar medium through stellar winds and radiation; 4) they leave, as remnant, exotic objects like neutron stars and black holes. Massive Population III Stars could play an important role in Cosmology because they contribute to 1) the reionization of the universe ad $z>5$, 2) the production of massive black holes that could have been the progenitors of active galactic nuclei, 3) the pregalactic metal enrichment. Finally Massive Stars play an important role in the field of γ-ray astrophysics because 1) they are responsible for the production of some long-lived γ-ray emitter nuclei as $^{26}$Al, $^{56}$Co, $^{57}$Co, $^{44}$Ti and $^{60}$Fe, and 2) they are likely connected to the Gamma Ray Bursts. As a consequence, the understanding of these stars, i.e., their presupernova evolution, their explosion as supernovae and, especially, their nucleosynthesis, is crucial for the interpretation of many astrophysical objects.

In this paper we present the basic properties of the yields of our latest presupernova evolution and explosion of massive stars, in the range between 11 and 120 $M_\odot$ having an initial solar and zero metallicity. A more detailed discussion of these models will be presented in a forthcoming paper.

# HYDROSTATIC AND HYDRODYNAMIC CODES

The results presented in this paper are based on a new set of presupernova models and explosions of solar and zero metallicity stars in the mass range between 11 and 120 $M_\odot$, covering therefore the full range of masses that are expected to give rise to Type II/Ib/Ic supernovae as well as those contributing to the Wolf-Rayet populations. All these models, that will be presented shortly in a forthcoming paper [1], have been computed by means of the more recent version (5.050218) of the FRANEC (Frascati Raphson Newton Evolutionary Code), whose main differences with the respect to the previous versions [2] are the following: 1) the time dependent mixing is taken into account by means of a classical diffusion equation [3]; 2) the equation for the convecitive mixing and the ones describing the chemical evolution of the matter due to the nuclear burning are coupled together and solved simultaneously; 3) a moderate amount of overshooting (0.2 $H_p$) is assumed during core H burning; 4) mass loss is taken into account following the prescription of [4,5] for the blue supergiant phase ($T_{eff}$>12000 K), [6] for the red supergiant phase ($T_{eff}$<12000 K) and [7] during the Wolf-Rayet phase; 5) updated cross sections have been adopted whenever possible (see electronic references table in [1]).

The explosion is simulated by means of a piston of initial velocity $v_0$ located at ~1 $M_\odot$ in the presupernova model and moving along a balistic trajectory under the gravitational field of the compact remnant. The developement and the evolution of the shock wave that forms, is followed by means of a 1D PPM lagrangian hydro code [8]. The explosive nucleosynthesis is computed by using the same nuclear network adopted in the hydrostatic evolution. For each model, several hydro calculations have been performed by iterating on $v_0$ in order to obtain a given amount of $^{56}$Ni ejected and a corresponding final kinetic energy at the infinity. Since, at present, there is no self consistent model for the explosion of a core collapse supernova, the relation between the initial mass and the remnant mass is essentially unknown. However, observations seem to indicate that stars with mass $M_{MS} \leq 25$ $M_\odot$ form neutron stars producing ~ 0.1 $M_\odot$ of $^{56}$Ni while stars with mass $M_{MS} \geq 25$ $M_\odot$ form black hole producing either ~ $10^{-3}$ $M_\odot$ or ~ 0.1 $M_\odot$ of $^{56}$Ni depending on many factors among which rotation, stellar wind, magnetic fields, metallicity and binarity [9]. Therefore, guided by the observations, we choose two initial mass-remnant mass relations: the first one (*trend*) in which we assume that stars with mass $M_{MS} \leq 25$ $M_\odot$ produce 0.1 $M_\odot$ of $^{56}$Ni while stars more massive than this limit produce $10^{-3}$ $M_\odot$ of $^{56}$Ni, the second one (*flat*) in which we assume that all the core collapse supernovae are assumed to eject 0.1 $M_\odot$ of $^{56}$Ni, independently of the initial mass.

# THE SOLAR METALLICITY MODELS

The first set of models discussed in the present paper is the one computed with initial solar metallicity [10] and including masses in the range between 11 and 120 $M_\odot$, namely, 11, 12, 13, 14, 15, 16, 17, 20, 25, 30, 35, 40, 60, 80 and 120 $M_\odot$. These models will be presented in more detail in a forthcoming paper. The first result worth to be mentioned is reported in Fig. 1, where the element production factors (PFs) averaged over a Salpeter Initial Mass Function ($dm/dn=km^{-2.35}$) are shown for the two chosen initial mass-remnant mass relations.

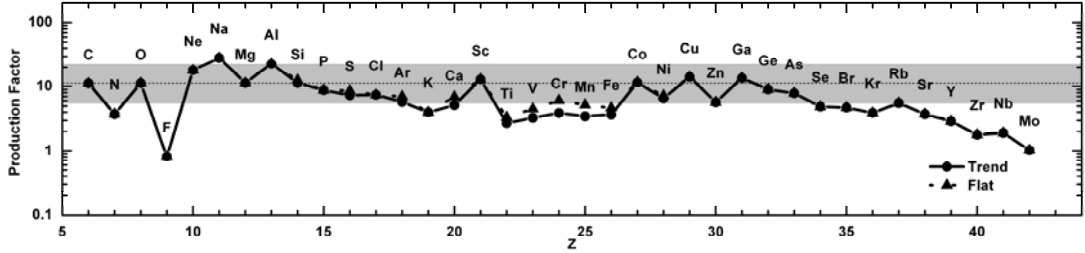

**FIGURE 1.** Element production factors averaged over a Salpeter Initial Mass Function for solar metallicty stars in the mass range 11- 120 $M_\odot$: the filled circles connected by the solid line refer to the *trend* case (see text); the filled triangles connected by the dotted line refer to the *flat* case. The horizontal dashed line refers to the production factor of O. All the nuclei whose production factors falls within a factor of 2 of the oxygen one (shaded area) are produced in scaled solar proportions.

In the reasonable assumption that the average metallicity grows continuously and slowly compared to the evolutionary timescales of the stars contributing to the global enrichment of the gas, it would be desirable that a generation of solar metallicity stars provides yields in roughly solar proportions or, in other words, that the PFs of the various isotopes remain essentially flat. Since oxygen is produced mainly by core collapse supernovae and it is also the most abundant element produced by these stars we use its production factor as the one that better represents the overall increase of the average metallicity and to verify whether or not the other nuclei follow its behavior. In particular, we assume that all the elements whose production factor falls within a factor of 2, taken as a suitable warning threshold, of the oxygen one are compatible with a flat distribution relative to oxygen, while those outside this compatibility range may potentially constitute a problem. The things worth noting in Fig. 1 are the following: (1) the only elements that vary substantially between the cases *trend* and *flat* are the iron peak ones, i.e., Ti, V, Cr, Mn and Fe; (2) the majority of the elements are produced in scaled solar proportions relative to O: the exceptions are N, F, K, the iron peak elements (Ti, V, Cr, Mn, Fe) and s-process elements heavier than Br. N and s-process elements above Br are under abundant, as expected, because these elements are mainly produced by intermediate mass stars that are not included in the mass interval analyzed in this paper. The underproduction of both F and K is mainly due to the lack, in these calculations, of the neutrino induced reactions [11]. The iron peak elements significantly depend on the adopted mass cut, i.e., the initial mass-remnant mass relation. Indeed, a changing from the *trend* to the *flat* case lead to an increase of the yields of these elements, pushing them to a closer scaled solar distribution and hence leaving less room for the SNIa contribution because of the larger amount of Fe ($^{56}$Ni) produced.

Figure 2 shows the element production factors averaged over a Salpeter IMF in the *trend* case, with two different upper mass limits, i.e., $M_{top}=35\ M_\odot$ and $M_{top}=120\ M_\odot$. Interestingly, Fig. 2 shows that the PFs of all the elements from N to Ca are almost independent on the upper mass limit, i.e., the PFs of all these elements are quite similar in all the models with initial masses in the range 35-120 $M_\odot$. The PFs of the iron peak elements increase by reducing $M_{top}$, as expected, because no iron is produced in stars more massive than 25 $M_\odot$ in the *trend* case.

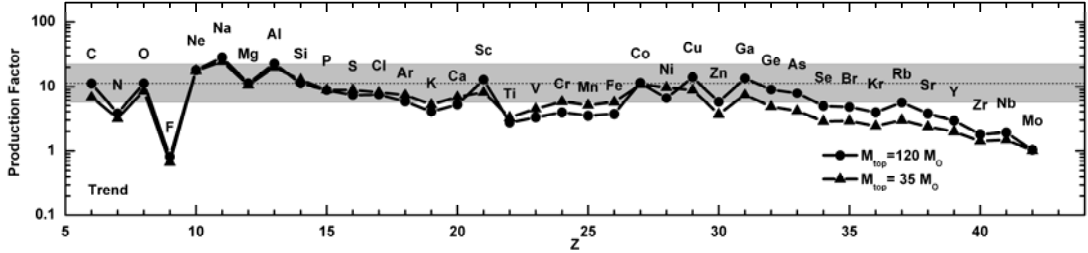

**FIGURE 2.** Element production factors averaged over a Salpeter Initial Mass Function for solar metallicty models in the *trend* case with two choices of the upper mass limit $M_{top}$ of the IMF: the filled circles connected by the solid line refer to the (standard) case in which $M_{top}$=120 $M_\odot$; the filled triangles connected by the dotted line refer to the case in which $M_{top}$=35 $M_\odot$. The horizontal dashed line refers to the production factor of O in the standard case.

The PFs of C and of the s-process elements decrease by reducing $M_{top}$ as a consequence of the substantial production of these elements in stars more massive than 35 $M_\odot$, hence the inclusion of stars in the mass range 35-120 $M_\odot$ changes the relative scaling of C and of the s-process relative to all the other elements (in particular O).

## THE ZERO METALLICITY MODELS

The set of zero metallicity models includes masses in the range between 13 and 80 $M_\odot$, namely, 13, 15, 20, 25, 30, 35, 50 and 80 $M_\odot$. At variance with the solar metallicity case, these models have been computed without mass loss and without overshooting and will be presented in more details in a forthcoming paper. The first interesting result to be mentioned is reported in Fig. 3 that shows the element production factors for all the computed models in the *trend* case. A large spread of the PFs of all the elements with Z<14 is clearly shown in Fig. 3. In particular a large primary N production is obtained in stars in the mass range between 25 and 35 $M_\odot$; such a primary production is connected, in these stars, with the ingestion of protons by the He convective shell that penetrates into the overlying H rich layer. Indeed, the protons ingested by the He convective shell activate the CNO cycle at high temperatures, typical of He burning, leading to a substantial production of N. Such a partial mixing between the He convective shell and the overlying H rich mantle is a rather common feature in stellar models of initial zero metallicity [11, 12, 13] because of the low entropy barrier present at the H-He interface in these stars. At variance with the lightest elements, the intermediate mass elements (14 ≤ Z ≤ 21) show PFs almost independent on the initial mass.

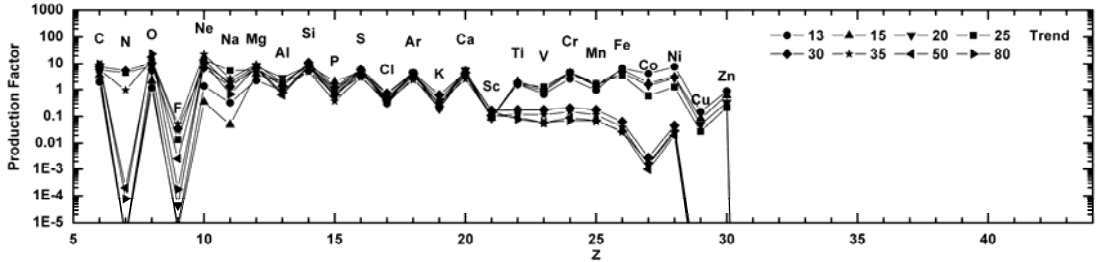

**FIGURE 3.** Production factors of all the elements obtained in the *trend* case for zero metallicity models: the symbols refer to the eight computed models as shown in the legend in the upper right corner.

The PFs of the iron peak elements, as expected, are larger for models producing more iron, i.e., stars below 25 $M_\odot$. As a consequence, such a behavior is strongly dependent on the initial mass-remnant mass relation. A last feature worth to be mentioned is that there is a cutoff in the PFs at the level of Zn, i.e., no elements heavier than Zn are produced in zero metallicity massive stars. This has the obvious consequence that the observed abundances of elements above Zn in very metal poor stars must be attributed to stars (or, in general, to processes) outside the range presently discussed [15].

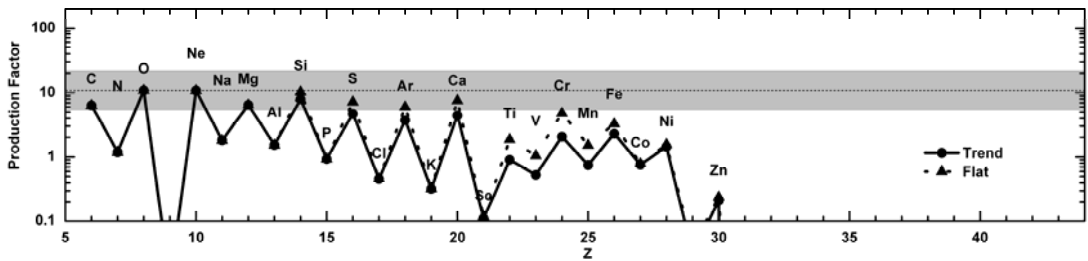

**FIGURE 4.** Element production factors averaged over a Salpeter Initial Mass Function for zero metallicity models: the filled circles connected by the solid line refer to the *trend* case (see text); the filled triangles connected by the dotted line refer to the *flat* case. The horizontal dashed line refers to the production factor of O. All the nuclei whose production factors falls within a factor of 2 of the oxygen one (shaded area) are produced in scaled solar proportions.

Although the primordial Initial Mass Function is still presently unknown, it is interesting to integrate the element yields of all the models over a Salpeter IMF. Figure 4 shows the element PFs averaged over a Salpeter IMF in the *trend* and *flat* cases. The first thing worth to be noted is that, as for the solar metallicity models, the PFs of all the elements are essentially independent on the initial mass-remnant mass relation, except those of the iron peak elements. The second interesting feature is the well known *odd-even effect*, i.e., the large difference between the PFs of the odd (from

N to Sc) and the even nuclei (from C to Ca). In particular, the elements C, Ne, Mg and Si are produced in almost scaled solar proportions relative to O while S, Ar and Ca are deficient by a factor of ~ 2. The odd elements, from N to Sc, are underproduced by a factor of 10 to 100 relative to O. The iron peak elements are deficient by a factor of about 10, although their production factors depend on the initial mass-remnant mass relation, i.e., they increase by changing from the *trend* to the *flat* case. As already shown in Fig. 3, no production of elements beyond Zn is obtained in these models.

## ACKNOWLEDGMENTS

Marco Limongi warmly thanks Ken'ichi Nomoto and the Organizing Committee for their financial support necessary to attend the OMEG05 meeting. Marco Limongi and Tatiana Grilli also thank Ken'ichi Nomoto, Keiichi Maeda, Nozomu Tominaga and Masaomi Tanaka for their very kind hospitality and support during their visit in Tokyo.